\documentclass{ifacconf}
\usepackage[normalem]{ulem}
\usepackage{color}
\usepackage{todonotes}
\usepackage{cite}
\usepackage{amsmath,amssymb,amsfonts}
\usepackage{graphicx}
\usepackage{mathtools} 
\usepackage{graphicx}      
\usepackage{natbib}        
\newcommand{\norm}[1]{\left\lVert#1\right\rVert}

\newtheorem{proposition}{Proposition}
\newtheorem{remark}{Remark}
\newtheorem{theorem}{Theorem}

\begin{document}
\begin{frontmatter}
	
	\title{Tuning Rules for a Class of Port-Hamiltonian Mechanical Systems}
	\author{Carmen Chan-Zheng,\thanksref{footnoteinfo2}}
	\author{Pablo Borja, and} 
	\author{Jacquelien M.A. Scherpen}
	
	\address{Jan C. Willems Center for Systems and Control.\\ Engineering and Technology Institute Groningen (ENTEG)\\University of Groningen, Nijenborgh 4, 9747 AG Groningen,\\ The Netherlands\\ (email:\{c.chan.zheng, l.p.borja.rosales, j.m.a.scherpen\}@rug.nl).}
	\thanks[footnoteinfo2]{The work of Carmen Chan-Zheng is supported by the University of Costa Rica (UCR).}
	\begin{abstract}  
		In this extended abstract, we propose a tuning approach for nonlinear mechanical systems to modify the behavior of the closed-loop system, where we are particularly interested in attenuating oscillations from the transient response. Towards this end, we inject damping into the system, and we provide two tuning methods to select the gains that are appropriate for our purposes.  Furthermore, we apply these tuning rules to a 2DoF planar manipulator and present its simulation results. 
	\end{abstract}
	
	
\end{frontmatter}
\section{Introduction}
The vast majority of the control methods for nonlinear physical systems often lack tuning guidelines since their main objective is to stabilize the closed-loop system. Nevertheless, from a practical implementation perspective, the need for these guidelines is important, sometimes even crucial. The current control techniques may neither be sufficient nor adequate as they are not necessarily designed to ensure that the closed-loop system has a prescribed performance.

In this work, we propose an approach to tune the gains for a particular class of physical systems, namely, fully actuated nonlinear mechanical systems. To our knowledge, only a few works address tuning guidelines for nonlinear systems. For instance, in \cite{jeltsema2004tuning} propose criteria based on Brayton-Moser equations for tuning switched-mode power converters.  Also, \cite{dirksz2013tuning} show that tuning the initial conditions improves the transient response of mechanical systems.

Customarily, damping injection is used to achieve asymptotic stability of the equilibrium point. However, in this work, we study the possibility of exploiting the damping injection to ensure the asymptotic stability property of the equilibrium and, additionally, a desired performance in the transient response of the closed-loop system. To this end, we assume that the open-loop nonlinear system has a stable equilibrium at the desired operating point. Then, we add a proportional controller in which we tune the gains later by using our methodology. Next, we analyze the closed-loop system with the help of its linearization. We look for a particular structure of a class of saddle point matrices, which can be found by a similarity transformation. After obtaining the required form, we apply a tuning rule to obtain either a ``no-overshoot'' response or a response with a particular damping ratio. 

Additionally, we provide a characterization of the region where the results are valid for the nonlinear system. This characterization is based on the estimation of the domain of attraction of the equilibrium point, see \cite{khalil2013nonlinear}.
	
\textbf{Caveat}: This work is accepted as a Late-Breaking result at the IFAC World Congress 2020.

\section{Preliminaries}
Consider the following linear system
\begin{equation}\label{sys}
\frac{d}{dt}\begin{bmatrix} \tilde{x}\\ \tilde{y}\end{bmatrix}=-\mathcal{N}\begin{bmatrix}\tilde{x}\\ \tilde{y}\end{bmatrix}, \quad \mathcal{N}:=\begin{bmatrix}X&Z^\top\\-Z&Y\end{bmatrix} 
\end{equation}
where  $X=X^\top \in \mathbb{R}^{n\times n}$ is positive definite, $Z \in \mathbb{R}^{m\times n}$ has full rank, $Y=Y^\top \in \mathbb{R}^{m\times m}$ is positive semidefinite, $\tilde{x} \in \mathbb{R}^n$, $\tilde{y} \in \mathbb{R}^m$ and  $m\leq n$. The matrix $\mathcal{N}$ corresponds to a class of \textit{saddle point matrices}. A detailed analysis of the eigenvalues of $\mathcal{N}$ can be found in \cite{benzi2006eigenvalues}. 
The advantage of using this structure is that by studying the properties of the spectra of one of its submatrices reveals interesting spectral properties of $\mathcal{N}$. For example,  the spectrum of $\mathcal{N}$ turns out to be positive and real if the following theorem is satisfied (\cite{benzi2006eigenvalues}): 
\begin{theorem}\label{t1}
	Assume $Y=0_{m \times m}$. Denote with $\lambda \in \mathbb{C}$ an eigenvalue associated with $\mathcal{N}$ and $\eta:=[v;w]^\top$ its corresponding eigenvector where $ v\in \mathbb{R}^{n}$ and  $ w\in \mathbb{R}^{m}$. Then $\lambda$ is real if and only if:
	\begin{align}\label{t1eq}
	\Bigg(\frac{v^* X v}{v^* v}\Bigg)^2\geq 4 \frac{v^*(Z^\top Z)v}{v^* v}.
	\end{align}
\end{theorem}
\begin{remark}\label{r1}
		Inequality \eqref{t1eq} is obtained from the definition of $\lambda$:
		\begin{align}\label{eig}
		\lambda:=\frac{1}{2}\begin{pmatrix}
		\frac{v^*Xv}{v^*v}\pm\sqrt{\Big(\frac{v^*Xv}{v^*v}\Big)^2-4\frac{v^*Z^\top Z v}{v^*v}}
		\end{pmatrix},
		\end{align}
		thus, $\lambda$ is real if and only if the discriminant is nonnegative. 
\end{remark}	
Furthermore, if $\lambda$ is non-real, then a bound can be provided for the real part by the following theorem (\cite{benzi2006eigenvalues}): 
\begin{theorem}\label{t2}
	Assume $Y=0_{m \times m}$ and the same notation as Theorem \ref{t1}. If $Im(\lambda)\neq 0$ then 
	\begin{align}
	\begin{split}
	2Re(\lambda)=\frac{v^* X v}{v^* v}.
	\end{split}
	\end{align}
\end{theorem}
For the proofs of Theorem \ref{t1} and Theorem \ref{t2}, we refer the reader to \cite{benzi2006eigenvalues}. 
\section{Problem Formulation}
Consider the following fully actuated Port-Hamiltonian (PH) nonlinear mechanical system:
\begin{align}\label{phsys}
\begin{split}
\begin{bmatrix}
\dot{q}\\ \dot{p}
\end{bmatrix} &= \begin{bmatrix}
0_{n\times n}&I_n \\-I_n&-D(q,p)
\end{bmatrix}\begin{bmatrix}
\frac{\partial H}{\partial q}\\\frac{\partial H}{\partial p} 
\end{bmatrix}+ \begin{bmatrix}
0_{n\times n}\\I_n
\end{bmatrix}u,\\
y&=\frac{\partial H}{\partial p}, \quad
H(q,p)=\frac{1}{2}p^\top M^{-1}(q)p + V(q)
\end{split}
\end{align}
where $q,p \in \mathbb{R}^n$ are the vectors of generalized positions and momenta, respectively, $u\in \mathbb{R}^n$ is the control vector, $y\in \mathbb{R}^n$ is the output of the system, $D(q,p) \in \mathbb{R}^{n \times n}$ is the system's damping verifying $D(q,p)=D(q,p)^\top\geq0$,  $H: \mathbb{R}^{2n} \to \mathbb{R}$ is the system's Hamiltonian, $M(q) \in \mathbb{R}^{n \times n} $ is the mass-inertia matrix satisfying $M(q)=M^\top(q)>0$ and $V(q)\in \mathbb{R}$ is the potential energy function of the system, which has an isolated minimum at the desired configuration $q^* \in \mathbb{R}^n$. The point $(q^*,0)$ is an isolated minimum of $H(q,p)$, and 
\begin{align}\label{hdot}
\dot{H}(q,p)=-\left(\frac{\partial H}{\partial p}(q,p)\right)^\top D(q,p)\frac{\partial H}{\partial p}(q,p) \leq 0
\end{align}
for $u=0$.

Thus, $(q^*,0)$ is a stable equilibrium of the open-loop system. 

The \textit{control objective} is to find an input $u$ such that the system described in \eqref{phsys} has either a ``no-overshoot" response or a response with a prescribed damping ratio.  In the sequel, we describe such input and provide a methodology to tune the gains of the controller. Furthermore, we provide an estimation of the domain of attraction of the nonlinear system where these rules are valid.

\section{Tuning rules for nonlinear mechanical systems}
Consider the control input $u=-K_t y$ with $K_t \in \mathbb{R}^{n \times n}$ verifying $K_t=K_t^\top >0$. Then the closed-loop system becomes:
\begin{align}\label{phsyscl}
\begin{split}
&\begin{bmatrix}
\dot{q}\\[3pt]\dot{p}
\end{bmatrix} = \begin{bmatrix}
0_{n\times n}&I_n \\[3pt]-I_n&-D(q,p)-K_t
\end{bmatrix}\begin{bmatrix}
\frac{\partial H}{\partial q}\\[3pt]\frac{\partial H}{\partial p} 
\end{bmatrix}
\end{split}.
\end{align}
\begin{remark}
	Since $(D(q,p)+K_t)>0$, the equilibrium point remains stable with the addition of $K_t$. Furthermore, 
	\begin{align}
	\dot{H}(q,p)=-\left(\frac{\partial H}{\partial p}(q,p)\right)^\top (D(q,p)+K_t)\frac{\partial H}{\partial p}(q,p) < 0.
	\end{align}
	Thus, it is fully damped and the asymptotic stability of the equilibrium can be proven by invoking LaSalle's invariance principle (see \cite{khalil2013nonlinear}).
\end{remark}
To obtain the linearized dynamics of \eqref{phsyscl}, we introduce the following vectors:
\begin{align}
\begin{split}
\tilde{q}=q-q^*, \quad
\tilde{p}=p.
\end{split}
\end{align}
Then, the linearized system around the equilibrium point $(q^*,0)$ corresponds to:
\begin{align}\label{sys2}
\begin{split}
\begin{bmatrix}
\dot{\tilde{q}}\\ \dot{\tilde{p}}
\end{bmatrix} &= A\begin{bmatrix}
\tilde{q}\\\tilde{p}
\end{bmatrix}, \quad A:=\begin{bmatrix}
0_{n\times n}&M^{-1}_* \\-P&-RM^{-1}_*
\end{bmatrix}\\
\end{split}
\end{align}

where $M_*^{-1}:=\frac{\partial^2 H}{\partial p^2}(q^*,0)$, $P:=\frac{\partial^2 H}{\partial q^2}(q^*,0)$ and $R:=D(q,p)\big\rvert_{p=0}+K_t$. 

In this extended abstract, we are particularly interested in the class of saddle point matrices described by \eqref{sys} as this specific structure facilitates the analysis of the transient response of a dynamical system. The main benefit of this special form is that the slowest eigenvalue of $\mathcal{N}$, i.e. closest to the imaginary axis, can be calculated based on its submatrices as seen in Remark \ref{r1}. This eigenvalue has a particular importance in our analysis since the oscillation behavior for the transient response is mainly characterized by it. Therefore, by modifying the value of $X$, we can prescribe a desired transient oscillation behavior for $\mathcal{N}$. Then, to apply the abovementioned advantage to system \eqref{sys2}, first, we need to transform $A$ into that class of saddle point matrix, hence, we need to find a similarity transformation matrix $W\in \mathbb{R}^{2n \times 2n}$. To obtain $W$, first consider the matrix decompositions
\begin{align}
M^{-1}_*=\phi_M^\top \phi_M, \quad
P=\phi^\top_P \phi_P
\end{align}
where $\phi_M \in \mathbb{R}^{n \times n}$ and $\phi_P\in \mathbb{R}^{n \times n}$ are full rank matrices.
Next, define $W$ and new coordinate $z\in \mathbb{R}^{2n}$ such that
\begin{align}
W:=\begin{bmatrix}
0_{n\times n}&\phi_M\\\phi_P&0_{n\times n}
\end{bmatrix}, \quad z:=Wx.
\end{align}

Then, the similar matrix is computed as follows
\begin{align}
\mathcal{N}=-WAW^{-1}=
\begin{bmatrix}
	\phi_M R\phi_M^\top&\phi_M \phi_P^\top\\
	-\phi_P\phi_M^\top &0_{n\times n} 
	\end{bmatrix}.
\end{align}
Note that $\mathcal{N}$ has a saddle point form as in system \eqref{sys} with $X=\phi_M R \phi_M^\top$, $Z=\phi_P \phi_M^\top$ and $Y=0_{n\times n}$. Therefore, the linearized system in the recently introduced coordinates corresponds to
\begin{align}\label{sys3}
\dot{z}=	-\mathcal{N}z.
\end{align}

In the following subsections, we propose two tuning methods to select the gains of matrix $R$ to obtain a desired oscillation behavior of the transient response of the system \eqref{phsyscl}.

\subsection{The ``No-Overshoot'' response case}
The behavior of the oscillations of the transient response is determined by the complex-conjugated poles of the system. The peak value measured from the equilibrium point of this corresponding oscillations is defined as the maximum overshoot (see \cite{kulakowski2007dynamic}).  	
In this section, we provide a condition such that system \eqref{sys3} presents a ``no-overshoot'' response. In other words, the matrix $\mathcal{N}$ from system \eqref{sys3} must contain only real spectrum, that is, the inequality \eqref{t1eq} must be satisfied. First, note from \eqref{t1eq} that the \textit{Rayleigh quotients}  $\frac{v^*Xv}{v^*v}$ and $\frac{v^*Z^\top Z v}{v^*v}$ are bounded by
\begin{align}
	&\lambda_{min}(X)\leq \frac{v^*Xv}{v^*v}\leq \lambda_{max}(X)\label{b1}\\
	&\lambda_{min}(Z^\top Z)\leq \frac{v^*Z^\top Z v}{v^*v} \leq \lambda_{max}(Z^\top Z)\label{b2}.
\end{align}	
Then, by using the bounds \eqref{b1} and \eqref{b2}, we propose the following:
\begin{proposition}\label{p1}
	The spectrum of the system \eqref{sys2} is real if the following is satisfied:
	\begin{align}\label{p1eq}
	(\lambda_{min}(X))^2	\geq 4 \lambda_{max}(Z^\top Z). 
	\end{align}
\end{proposition}
\begin{remark}
		Note from \eqref{eig} that modifying the value of the minimum eigenvalue of $X$ will also modify the value of the slowest pole of system \eqref{sys3}. 
\end{remark}
\textit{Proof}: Note that
	\begin{align}\label{cond2}
	(\lambda_{min}(X))^2\leq \Bigg(\frac{v^*X v}{v^* v}\Bigg)^2 \leq (\lambda_{max}(X))^2,
	\end{align}
	and 
	\begin{align}\label{cond3}
	4 \lambda_{min}(Z^\top Z) \leq 4 \frac{v^*(Z^\top Z)v}{v^* v} \leq 4 \lambda_{max}(Z^\top Z),
	\end{align}
	hence, if \eqref{p1eq} holds, then inequality \eqref{t1eq} from Theorem \ref{t1} is satisfied since
	\begin{align}
 4 \frac{v^*(Z^\top Z)v}{v^* v}\leq 4 \lambda_{max}(Z^\top Z) \leq(\lambda_{min}(X))^2\leq 	\Bigg(\frac{v^*X v}{v^* v}\Bigg)^2 .
	\end{align}
$\square$

Proposition \ref{p1} is rather conservative, since it may give a larger value for the lower bound for $X$ than necessary. This may result in an over-damped response, i.e., a slower settling time. However, due to the mechanical system structure, we can provide a smaller bound by using:
\begin{align}\label{p1c}
(\lambda_{min}(R))^2	\geq 4 \lambda_{max}(M_*)\lambda_{max}(P).
\end{align}
\begin{remark}
	By choosing $\lambda_{min}(R)$, such that $(\lambda_{min}(R))^2=4\lambda_{max}(M_*)\lambda_{max}(P)$ results in a critically damped response.
\end{remark}

\subsection{The underdamped response case}
The tuning rule proposed in \eqref{p1c} might be restrictive for some applications that need a faster settling time. However, this is usually achieved at the expense of a transient response with overshoot and oscillations. If this performance is acceptable, we can propose a tuning rule to improve the settling time. This is done next.

Denote with $\lambda_p, \bar{\lambda}_p \in \mathbb{C}$ the slowest pair of complex conjugated eigenvalues of $\mathcal{N}$, then, one of the parameters used to characterize the transient response's oscillations is the damping ratio which is defined as 
\begin{align}\label{dampratio}
\zeta_p:=\frac{|Re(\lambda_p)|}{\sqrt{Re(\lambda_p)^2+Im(\lambda_p)^2}}
\end{align}
where $0\leq \zeta_p\leq 1$.

The transient oscillations behavior is mostly influenced by this pair of complex conjugated poles (dominant-pair) (\cite{kulakowski2007dynamic}).  

In this extended abstract, we propose a tuning rule, based on Theorem \ref{t2}, in which we can modify the damping ratio of the dominant-pair, and hence, influence the performance of the system \eqref{phsyscl}. 

Assuming that system \eqref{sys3} has at least a pair of complex conjugated poles, then, we propose the following:
\begin{proposition}\label{p2}
	
	Let $\lambda \in \mathbb{C}$ a nonreal eigenvalue of $\mathcal{N}$ and $[v;w]^\top$ its corresponding eigenvector with $v\in \mathbb{R}^n$ and $w\in \mathbb{R}^n$. Then, the following relations hold: 
	\begin{subequations}
		\begin{align}
		\frac{v^*Xv}{v^*v}&=2Re(\lambda)\label{p2a}\\
		\frac{v^*Z^\top Z v}{v^*v}&=Re(\lambda)^2+Im(\lambda)^2.\label{p2b}
		\end{align}
	\end{subequations}
\end{proposition}
\textit{Proof:} The proof of \eqref{p2a} can be found in \cite{benzi2006eigenvalues}, the sketch of this proof reads as follows. Given the particular structure of  $\mathcal{N}$, we have the following property:
\begin{align}\label{normprop}
\norm{v}=\norm{w}.
\end{align}
Then, consider the following equality:
\begin{align}\label{eigpro}
\begin{bmatrix}X&Z^\top\\-Z&0\end{bmatrix}\begin{bmatrix}v\\w\end{bmatrix}= \lambda \begin{bmatrix}v\\w\end{bmatrix}.
\end{align}
By rearranging the system of equations in \eqref{eigpro} and using \eqref{normprop}, we obtain the following:
\begin{align}\label{lambdaeq}
\lambda^2-\frac{ v^* X v}{v^* v}\lambda+ \frac{v^* Z^\top Z v}{v^* v}=0.
\end{align}
The solution of \eqref{lambdaeq} is given by \eqref{eig}. Since $\lambda \in \mathbb{C}$, it follows that $Re(\lambda)=\frac{1}{2}\frac{v^*Xv}{v^*v}$.

To proof the equality \eqref{p2b}, note that substituting \eqref{p2a} in \eqref{lambdaeq} we have that

\begin{subequations}\label{modulus}
\begin{align}
&\lambda^2-2Re(\lambda)\lambda+\frac{v^*Z^\top Z v}{v^*v}=0\nonumber\\
&\implies \frac{v^*Z^\top Z v}{v^*v}=-(iIm(\lambda)+Re(\lambda))^2\nonumber\\
& \qquad \qquad  \qquad  \qquad  \qquad +2Re(\lambda)(iIm(\lambda)+Re(\lambda)) \nonumber\\
&\implies \frac{v^*Z^\top Z v}{v^*v}=(Re(\lambda))^2+(Im(\lambda))^2	\nonumber,
\end{align}
\end{subequations}
this completes the proof. $\square$

The relations described in Proposition \ref{p2} suggest that the damping ratio of $\lambda$ can be defined as:
\begin{align}
	\zeta := \frac{1}{2}\frac{v^*Xv}{v^*v}\Bigg(\sqrt{\frac{v^*Z^\top Z v}{v^*v}}\Bigg)^{-1}.
\end{align}
Therefore, we can modify the lower bound of the system's damping ratio with the following tuning rule:
\begin{align}\label{p2c}
\zeta^2\geq\frac{1}{4}\frac{\lambda_{min}(R)^2}{\lambda_{max}(M_*)\lambda_{max}(P)}
\end{align}	
where $0\leq\zeta\leq 1$.
\begin{remark}
Note that the relation \eqref{p2c} is monotonically increasing, that is, the smallest eigenvalue of $Z^\top Z$ influences directly the smallest eigenvalue of $X$.
\end{remark}

\subsection{Domain of attraction}
The proposed tuning rules are based on the linearized system, i.e., these rules are valid in a neighborhood of the nonlinear system around the equilibrium point $(q^*,0)$. Therefore, in this section we provide an estimate of how far the trajectories of the nonlinear system can be from the equilibrium point and still converge. In general, estimating the domain of attraction by using $H(q,p)$ as the Lyapunov candidate is a difficult task (see \cite{kloiber2012estimating} for a estimation algorithm).  For this extended abstract, we use a simpler method to estimate the domain as explained in \cite{khalil2013nonlinear}. 

Consider the quadratic Lyapunov candidate $V(x)=x^\top Px$ for the system \eqref{phsyscl} with $x:=\begin{bmatrix}q&p\end{bmatrix}^\top$ and $P \in \mathbb{R}^{2n \times 2n}$ verifying $P=P^\top>0$. For a given matrix $Q \in \mathbb{R}^{2n \times 2n}$ satisfying $Q=Q^\top >0$, select $P$ such that $Q=-(PA+A^\top P)$, where $A$ is defined as in \eqref{sys2} and is Hurwitz.\footnote{The matrix $Q$ definition is the well-known \textit{Lyapunov equation}} Next, rewrite \eqref{phsyscl} as
\begin{align} \label{phsyscl1}
\dot{x}=Ax+f_r(x)
\end{align}
where  $f_r(x) \in \mathbb{R}^{2n}$ is the remainder such that
\begin{align}
\frac{\norm{f_r(x)}}{\norm{x}} \to 0 \quad \textit{as} \quad \norm{x}\to 0.
\end{align}
Hence, for any $\gamma>0$, there exists $\rho>0$ such that
\begin{align}
\norm{f_r(x)}<\gamma \norm{x},\quad \forall \norm{x}<\rho.
\end{align}	
Then, it follows from some non-trivial calculations that
\begin{align}\label{limit}
\dot{V}&\leq -(\lambda_{min}(Q)-2\gamma\norm{P})\norm{x}^2<0,
\end{align}
for $\gamma<\frac{\lambda_{min}(Q)}{2\norm{P}}$. Therefore, $(q^*,0)$ is asymptotically stable in the domain $D:=\{x\in \mathbb{R}^n | \norm{x}<\rho\}$. Moreover, we can show again by some non-trivial steps that 
\begin{align}\label{explimits}
\lambda_{min}(P)\norm{x}^2\leq V(x)\leq \lambda_{max}(P)\norm{x}^2,
\end{align}
and thus, the equilibrium point is exponentially stable, see Theorem 4.10 in \cite{khalil2013nonlinear}. Furthermore, an estimation of the domain of attraction is given by the subset
\begin{align}
\Omega_c=\{x\in \mathbb{R}^n| V(x)\leq c\},
\end{align}
where $c<\lambda_{min}(P) \rho^2$.

\section{Simulations}
In this section, we apply the proposed controller to a 2 DoF planar manipulator (See \cite{quanser} for reference manual). The closed-loop model of the system, obtained from \cite{dirksz2011power}, along with our controller $u$ is shown below 
\begin{align}\label{phsysclplanar}
\begin{split}
&\begin{bmatrix}
\dot{q}\\ \dot{p}
\end{bmatrix} = \begin{bmatrix}
0_{2\times 2}&I_2 \\-I_2&-K_d
\end{bmatrix}\begin{bmatrix}
\frac{\partial H}{\partial q}\\\frac{\partial H}{\partial p} 
\end{bmatrix}+\begin{bmatrix}0_{2\times 2}\\I_2\end{bmatrix}u,\\
&H(q,p)=\frac{1}{2}p^\top M^{-1}(q)p + \frac{1}{2}(q-q^*)^\top K_p (q-q^*).
\end{split}
\end{align}
The mass-inertia matrix is given by
\begin{align}
M(q)=\begin{bmatrix}
a_1+a_2+2b\cos(q_2)&a_2+b\cos(q_2)\\
a_2+b\cos(q_2)&a_2
\end{bmatrix}
\end{align}
with $ a_1:=m_1r_1^2+m_2l_1^2+I_1$, $a_2:=m_2r_2^2+I_2$ and $b:=m_2l_1r_2$. The parameter $l_i$ is the link length, $m_i$ is the link mass, $r_i$ is the distance of the link to the center of the mass and $I_i$ is the moment of inertia with $i=1,2$. These parameters are obtained from \cite{dirksz2011power} and are shown along with the matrices $K_p$ and $K_d$ in Table \ref{simt}. 
\begin{remark}
	In this example, the 2DoF planar manipulator is previously stabilized with a energy shaping plus damping injection controller.
\end{remark}

\begin{table}[t]
	\caption{Simulation parameter values}\label{simt}
	\centering
	\begin{tabular}{lr}
		\hline
		\multicolumn{1}{c}{Parameter} & \multicolumn{1}{c}{Value} \\ \hline
		$m_1/m_2$                       & 0.5/1                     \\
		$I_1/I_2$                       & 0.01/0.01                 \\
		$r_1/r_2$                       & 0.2/0.25                  \\
		$l_1/l_2$                       & 0.343/0.275               \\
		$K_p$                           & diag(20,20)               \\
		$K_d$                           & diag(1,1)                 \\ \hline
	\end{tabular}
\end{table}

\begin{figure}[t]
	\centerline{\includegraphics[width=0.5\textwidth]{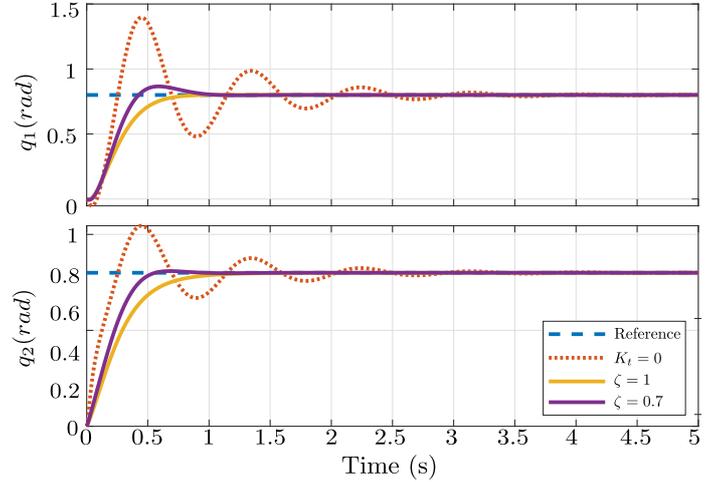}}
	\caption{Trajectories for angular position. Initial conditions $(q,p)=(0_2,0_2)$}\label{pos}
\end{figure}

In the simulations, our objective is to stabilize the 2DoF planar manipulator's joints to an angle of $0.8$ $rad$. The simulation of the trajectories for the angular position is shown in Fig. \ref{pos}. When $K_t=0$, it can be seen that the transient response for the angular position contains oscillations where it was attenuated when applying any of the tuning rules.

For the ``no-overshoot'' scenario ($\zeta=1$), we proceeded to apply the tuning rule \eqref{p1c}. The matrix $K_t$ is selected such that $(\lambda_{min}(X))^2 = 4 \lambda_{min}(Z^\top Z)$. Note that this value assignment corresponds to a critical damped response. A value lower than $4 \lambda_{min}(Z^\top Z)$ will result in a response with overshoot as the matrix $\mathcal{N}$ will contain eigenvalues with nonzero imaginary parts. As shown in Fig. \ref{pos}, both trajectories do not present any oscillations in their transient response.

For the underdamped scenario ($\zeta=0.7$), we apply the tuning rule \eqref{p2c}. The matrix $K_t$ is selected such that the dominant-pair has a damping ratio of $\zeta=0.7$. Fig. \ref{pos} shows the improved response where it has a faster settling time in comparison with the responses of $K_t=0$ and $\zeta=1$.

\section{Conclusions and future research}
Initial results have shown that transforming the PH structure into other coordinates reveals interesting spectral properties which can be used to improve the transient response for the nonlinear mechanical systems. As illustrated in the simulations, the proposed tuning rules are able to prescribe a desired performance in terms of the oscillation behavior of the closed-loop system. 

Possible future research includes providing a less conservative estimation of the region where the equilibrium of the nonlinear closed-loop system is exponentially stable. Additionally, we aim to extend this methodology to underactuated mechanical systems.

\bibliography{ifacconf}             

\end{document}